\documentstyle[preprint,aps,epsf]{revtex}
\newcommand{\del}{\mbox{$\Delta$}}
\newcommand{\arr}{\mbox{$\rightarrow$}}
\newcommand{\pinphin}{\mbox{$\pi N \arr N \phi \/$ }}
\newcommand{\pidelnphi}{\mbox{$\pi \del \arr N \phi \/$ }}
\newcommand{\nnnnphi}{\mbox{$N N \arr N N \phi \/$ }}
\newcommand{\ndelnnphi}{\mbox{$N \del \arr N N \phi \/$ }}
\newcommand{\deldelnnphi}{\mbox{$\del \del \arr N N \phi \/$ }}

\begin{document}

\title{Phi meson production cross sections in pion-baryon and
baryon-baryon interactions}

\author{W. S. Chung$^1$ \and  G. Q. Li$^{1,2}$ and C. M. Ko$^{1}$}

\address{$^1$Department of Physics and Cyclotron Institute,
Texas A\&M University, \\
College Station, Texas 77843, U.S.A.\\
$^2$Department of Physics, State University of New York at Stony Brook,\\
Stony Brook, New York 11794, U.S.A.}

\maketitle
 
\begin{abstract}
Phi meson production in pion-baryon and baryon-baryon
interactions near the production threshold is studied in
a boson-exchange model. Parameters in this model
are either taken from the Bonn model for the nucleon-nucleon
potential or determined from the measured phi meson decay width and
the experimental data for the reactions
$\pi^-p\arr \phi n$ and $pp\arr pp\phi$. The isospin-averaged cross
sections for the reactions $\pi N\arr \phi N$, $\pi \Delta
\arr \phi N$, $NN\arr NN\phi$, which are needed in transport 
models to study phi meson production in heavy-ion collisions at
subthreshold energies, have been evaluated. For the reactions $N\Delta\arr
NN\phi$ and $\Delta\Delta\arr NN\phi$, which become singular when the
exchanged pion is on-shell, the Peierls method is used to calculate their
cross sections. We find that the cross section for the reaction
$\pi N\arr\phi N$ is much larger than the other cross
sections.
\end{abstract}

\newpage
 
The production of phi mesons, which are pure ${\bar ss}$ states, in hadronic
interactions is usually suppressed due to the Okubo-Zweig-Iizuka (OZI)
rule \cite{hols92}.  An enhanced production of phi mesons in
ultra-relativistic heavy-ion collisions has thus been suggested as a
possible signal for the deconfinement phase transition of hadrons to 
a quark-gluon plasma \cite{shor85}, where the strangeness production
rate has been shown to be order-of-magnitude larger than in
a free hadronic gas. However, the phi meson properties might be changed
in hot dense matter due to the partial restoration of chiral symmetry.
Indeed, studies based on QCD sum rules \cite{hat92,ko94},
hadronic models including vacuum polarization effects \cite{hat95}, and the
hidden gauge theory \cite{song96} have all shown that the phi meson mass
is reduced in medium. The change of phi meson properties can be studied
in dilepton spectra \cite{shur92} and may also be inferred
from the phi meson yield in heavy-ion collisions \cite{ko91}.

Phi meson production from heavy-ion collisions has been studied in
CERN experiments from the dimuon invariant mass spectra \cite{na38}.
It has been found that the ratio for $\phi/\omega$ in S+U collisions at
200 GeV/nucleon is enhanced by about a factor of 3 compared to that from
the proton-proton and p-U interactions at the same energy. The enhancement
can probably be accounted for by assuming that a quark-gluon plasma has been 
formed in the collisions. On the other hand, it can also be explained by 
hadronic scenarios \cite{ko91,rqmd,koch90}. In particular, it has been 
pointed out in Ref. \cite{ko91} that the reactions $K\Lambda\to\phi N$ and
$K\bar K\to\phi\rho$ become important if one takes into account the medium
effects on hadron masses in hot dense matter.
 
Phi mesons have also been measured recently in heavy-ion collisions at
AGS energies from the invariant mass distribution of kaon-antikaon pairs
\cite{ags}, and their mass and width are found to be consistent with
that in free space. This is not surprising as these kaon-antikaon pairs
are from the decay of phi mesons at freeze out when their properties are 
the same as in free space. If a phi meson decays in medium, the 
resulting kaon and antikaon would interact with nucleons, so their 
invariant mass is modified and can no longer be used to reconstruct 
the phi meson. 

Phi meson production from heavy-ion collisions at SIS/GSI energies
are being studied by the FOPI collaboration 
also through the $K^+K^-$ invariant mass distribution \cite{fopi}.
Their preliminary results indicate that the $\phi /K^-$ ratio
is about 10\%, and is thus similar to that observed at AGS energies.
This is quite surprising as the incident energies at GSI and AGS differ by
order of magnitude. In a previous study by us \cite{liko95a}, the phi meson 
yield in central Ni+Ni collisions at 2 GeV/nucleon has been studied via
its decay into dileptons. Including
phi meson production only from the kaon-antikaon annihilation channel,
the $\phi/K^-$ ratio was found to be about 1\%, which is much below the 
value of 10\% from the preliminary data of the FOPI collaboration  
\cite{fopi}. It is thus necessary to also consider phi meson 
production from other processes, such as the baryon-baryon 
and pion-baryon interactions.

There are a number of advantages to work at heavy-ion collisions
at SIS energies, which are below the threshold for strange particle production
from nucleon-nucleon interactions in free space.
First, it has been shown that subthreshold particle production in heavy ion 
collisions is sensitive to medium modifications of hadron properties 
\cite{cass90,mosel,koli96}.
Second, at these energies the reaction dynamics is relatively simple as
the colliding system consists mainly of nucleons,
delta resonances, and pions. In addition to kaon-antikaon annihilation as
considered in Ref. \cite{liko95a}, phi mesons can be produced from the
pion-nucleon, pion-delta, nucleon-nucleon, nucleon-delta,
and delta-delta collisions.  The cross sections for
these reactions are thus needed as inputs in the transport model.
Unfortunately, these cross sections are not known empirically near the
threshold, which are relevant for phi meson production in heavy ion
collisions at subthreshold energies. In this letter, we shall introduce 
a boson-exchange model to calculate these cross sections. Similar
models have been used previously to study pion \cite{eng}, kaon
\cite{wuko,laget,liko95b}, eta meson \cite{peters}, and dilepton
\cite{grei} production in nucleon-nucleon interactions in the same
energy region.

Specifically, we shall calculate the following cross sections:
\pinphin , \pidelnphi , \nnnnphi , \ndelnnphi , and \deldelnnphi .
In principle, one needs for heavy ion collisions
the inclusive cross sections such as $NN\rightarrow \phi X$ instead of
the exclusive cross sections such as \nnnnphi.
Since the energies at SIS are below the phi meson production threshold in the
nucleon-nucleon interaction, we expect that exclusive reactions such as
\nnnnphi, which have lower thresholds, are more important than the
ones with pions also in the final state, which
require additional energies and are thus suppressed. We note that 
the boson-exchange model can be
generalized to the case with one or two pions in the final state
using the approximation of resonance dominance, namely, replacing one
or both nucleons in the final state by the delta resonance.
Such studies will be pursued in the future.

The Feynman diagrams for \pinphin and \pidelnphi are shown in Fig. 1,
while those for \nnnnphi ,\ndelnnphi , and \deldelnnphi are shown in Fig. 2,
Fig. 3, and Fig. 4, respectively. For baryon-baryon interactions 
there is also the exchange diagram as the final nucleons are identical.
The Lagrangians for $\pi NN$, $\rho NN$,  $\rho N\Delta$ \cite{mach89},  
and $\pi\rho\phi$ \cite{meis88} interactions are well-known, 
and they are given by
\begin{eqnarray}
{\cal L} _{\pi NN}& =& - {f_{\pi NN}\over m_\pi} {\bar \psi} \gamma ^5 \gamma
^\mu\vec\tau \psi\cdot \partial _\mu \vec\pi,\\
{\cal L} _{\rho NN}& =& -g_{\rho NN} {\bar \psi} \gamma ^\mu\vec\tau\psi
\cdot\vec\rho_\mu - {f_{\rho NN}\over 4m_N} {\bar \psi} \sigma ^{\mu\nu}
\vec\tau\psi\cdot (\partial _\mu\vec\rho_\nu -\partial _\nu\vec\rho_\mu),\\
{\cal L} _{\rho N \Delta}& =& i{f_{\rho N\Delta}\over m_\rho} {\bar \psi}
\gamma ^5 \gamma ^\mu\vec{\bbox T} \psi ^\nu\cdot(\partial_\mu\vec\rho_\nu
-\partial_\nu\vec\rho_\mu) + h.c.,\\
{\cal L} _{\pi\rho \phi}& =& {f_{\pi\rho\phi}\over m_\phi} 
\varepsilon ^{\mu\nu\alpha\beta}
\partial_\mu\vec\rho_\nu\partial_\alpha\phi_\beta\cdot\vec\pi.
\end{eqnarray}
In the above, $\psi$ is the nucleon field with mass $m_N$; $\psi _\mu$ is the
Rarita-Schwinger field for the spin-3/2 $\Delta$ resonance; and
$\pi$, $\rho_\mu$, $\phi_\mu$ are meson fields with masses $m_\pi$, 
$m_\rho$, $m_\phi$, respectively. 
The isospin matrices, Dirac $\gamma$-matrices and 
Levi-Civita tensor are denoted by $\vec\tau$, $\gamma^\mu$ and 
$\varepsilon ^{\mu\nu\alpha\beta}$, respectively.

The matrix element for the reaction \pinphin includes
both the vector and tensor contributions and is given by
\begin{eqnarray}
M&=&{ig_{\rho NN} f_{\pi\rho\phi} \over m_\phi (t-m_\rho ^2)}
\Big[{\bar u}(p_f) \gamma^\nu u(p_i)\Big] \varepsilon _{\alpha\beta\gamma\nu}
q^\alpha \varepsilon ^\beta k^\gamma \nonumber\\
&-&{f_{\rho NN}f_{\pi\rho\phi} \over 2m_Nm_\phi (t-m_\rho^2)}
\Big[{\bar u}(p_f)\sigma ^{\rho\nu} k_\rho u(p_i)\Big]
\varepsilon_{\alpha\beta\gamma\nu}q^\alpha \varepsilon ^\beta k^\gamma.
\end{eqnarray}
In the above, $u$ and ${\bar u}$ are Dirac spinors of the initial
and final nucleons with momenta $p_i$ and $p_f$, respectively. The momentum
and polarization of the phi meson are denoted by $q$ and
${\bbox\varepsilon}$, respectively, while $t$ is the square of 
the four-momentum $k$ of the exchanged rho meson.

The differential cross section is then
\begin{eqnarray}
{d\sigma \over dt}={1\over 64 \pi s} {1\over |{\bf p_\pi}|^2}
|M|^2,
\end{eqnarray}
where ${\bf p_\pi}$ and $\sqrt s$ are, respectively, the pion three-momentum
and the total energy in the center-of-mass frame of the pion-nucleon system.
In the above equation, the sum over final spins and average over initial 
spins are implied. We note that in evaluating $|M|^2$ in the above
equation and also that for other reactions discussed below,
the program Form (Version 1.0) \cite{verm} has been extensively used. 

The coupling constants ($f_{\pi NN}$, $g_{\rho NN}$, $f_{\rho NN}$,
$f_{\rho N\Delta}$) and cut-off parameters in ${\cal L}_{\pi NN}$,
${\cal L}_{\rho NN}$, and ${\cal L}_{\rho N\Delta}$ are taken from the Bonn
one-boson-exchange model (Model II) as listed in Table B.1 of Ref. 
\cite{mach89}. From the measured width $\Gamma _{\phi
\rightarrow \pi \rho}\approx 0.6$ MeV, the coupling constant
$f_{\pi\rho\phi}\approx 1.04 $ is determined. In addition to the
form factor at the $\rho NN$ vertex which is taken from the Bonn
model, we also introduce a monopole form factor with a cut-off parameter
$\Lambda_{\pi\rho\phi}^\rho$ at the $\pi\rho\phi$ vertex.
By fitting to the available experimental
data for the reaction $\pi^-p\rightarrow\phi n$ \cite{data1},
we obtain $\Lambda_{\pi\rho\phi}^\rho=1.2$ GeV.
The comparison of the calculated cross section (solid curve) 
with the data (open circles)
is shown in Fig. 5. We see that the dominant contribution comes from
the tensor $\rho NN$ coupling (long-dashed curve) as that from 
the vector $\rho NN$ coupling (short-dashed curve) and the interference 
term (dot-dashed curve) are small. In the same figure, we also show the
parameterization introduced by Sibirtsev \cite{sib} (dotted curve). Although 
both are similar at energies where data exist, they differ appreciably
near the threshold.

Having determined the cut-off parameter, we can calculate the
isospin-averaged cross sections for \pinphin and \pidelnphi 
without introducing further free parameters. In our study the $\Delta$
particle is treated by the usual Rarita-Schwinger formalism,
with its spin projection operator given by
\begin{eqnarray}
 {\it P}^{\mu \nu} = -(\rlap/P + m_\Delta ) 
       \left( g^{\mu \nu} 
     -{1 \over 3} \gamma^{\mu} \gamma^{\nu} 
     - {2 P^{\mu} P^{\nu} \over 3 m^2_\Delta } 
     + { P^{\mu} \gamma^{\nu} - P^{\nu} \gamma^{\mu} 
         \over 3 m_\Delta } \right).
\end{eqnarray}
The results are shown in Fig. 6. We see that the cross section
for \pinphin (solid curve) is about a factor of five larger than that of
\pidelnphi (dotted curve), mainly because of the tensor $\rho NN$ coupling
in the former reaction. 

The matrix element for \nnnnphi includes both
the direct and exchange terms and is given by
\begin{eqnarray}
M=M_d+M_e,
\end{eqnarray}
where
\begin{eqnarray}
M_d&=&-{f_{\pi NN}g_{\rho NN}f_{\pi\rho\phi}\over m_\pi m_\phi
(q_{1d}^2-m_\pi^2) (q_{2d}^2-m_\rho^2)}           
\Big[{\bar u}(p_3) \gamma _5 \rlap/{q_{1d}} u(p_1)\Big]
\Big[{\bar u}(p_4)\gamma^\beta u(p_2)\Big]
\varepsilon _{\mu\nu\alpha\beta} q^\mu\varepsilon_\nu q_{2d}^\alpha\nonumber\\
&-&{if_{\pi NN}f_{\rho NN}f_{\pi\rho\phi}\over 2m_N m_\pi m_\phi
(q_{1d}^2-m_\pi^2) (q_{2d}^2-m_\rho^2)}            
\Big[{\bar u}(p_3) \gamma _5 \rlap/{q_{1d}} u(p_1)\Big]
\Big[{\bar u}(p_4) \sigma _{\rho\beta} u(p_2)\Big] \nonumber\\
&\cdot&\varepsilon ^{\mu\nu\alpha\beta} q_\mu\varepsilon_\nu (q_{2d})_\alpha 
q_{2d}^\rho ,
\end{eqnarray}
and
\begin{eqnarray}
M_e&=&-{f_{\pi NN}g_{\rho NN}f_{\pi\rho\phi}\over m_\pi m_\phi
(q_{1e}^2-m_\pi^2) (q_{2e}^2-m_\rho^2)}            
\Big[{\bar u}(p_4) \gamma _5 \rlap/{q_{1e}} u(p_1)\Big]
\Big[{\bar u}(p_3)\gamma^\beta u(p_2)\Big]
\varepsilon _{\mu\nu\alpha\beta} q^\mu\varepsilon_\nu q_{2e}^\alpha\nonumber\\
&-&{if_{\pi NN}f_{\rho NN}f_{\pi\rho\phi}\over 2m_N m_\pi m_\phi
(q_{1e}^2-m_\pi^2) (q_{2e}^2-m_\rho^2)}            
\Big[{\bar u}(p_4) \gamma _5 \rlap/{q_{1e}} u(p_1)\Big]
\Big[{\bar u}(p_3) \sigma _{\rho\beta} u(p_2)\Big] \nonumber\\
&\cdot&\varepsilon ^{\mu\nu\alpha\beta} q_\mu\varepsilon_\nu (q_{2e})_\alpha 
q_{2e}^\rho .
\end{eqnarray}
In the above, $p_1$ and $p_2$ are the four-momenta of two initial
nucleons; $p_3$ and $p_4$ are those of final nucleons; and q and $\varepsilon$
are, respectively the momentum and polarization of the phi meson.
The momenta of the exchanged pion and rho meson are denoted, respectively,
by $q_{1d}=p_3-p_1$ and $q_{2d}=p_4-p_2$ for the direct term and
by $q_{1e}=p_4-p_1$ and $q_{2e}=p_3-p_2$ for the exchange term.

For the reaction $NN\to NN\phi$, in addition to the usual form factors
at the $\pi NN$ and $\rho NN$ vertices taken from the Bonn potential
model, two additional monopole from factors are
introduced at the $\pi\rho\phi$ vertex as both pion and rho meson
are off-shell. The one associated with the rho meson
is taken to be the same as in the reaction $\pi N\to N\phi$.
The other one associated with the pion introduces another
cut-off parameter $\Lambda_{\pi\rho\phi}^\pi$, and it can be determined by
fitting to the cross section for the reaction $pp\rightarrow pp\phi$.
Unfortunately, no experimental data are available for this reaction
near the threshold where the boson-exchange model is expected to be
appropriate.  Nevertheless, we choose
$\Lambda ^\pi_{\pi\rho\phi}$=0.95 GeV to fit the
experimental data at $p_{lab}\approx 10$ GeV \cite{data2,data3}, 
which is the lowest beam energy with experimental data available.
The resulting cross section is shown in Fig. 7 by the solid curve
together with the experimental data (open circle). 
It should be noted that $\Lambda_{\pi\rho\phi}^\rho$ and
$\Lambda ^\pi_{\pi\rho\phi}$ are both in the order of 1 GeV,
similar to typical values for cut-off parameters in the Bonn potential model.

The cross section for $pp\rightarrow pp\phi$ has also been
studied by Sibirtsev \cite{sib} based on an one-pion-exchange
model. In this study, both the off-shell feature of the pion and the
interference between the direct and exchange diagrams are neglected,
and the cross section for $pp\to pp\phi$ can thus be expressed in terms of the 
$\pi^-p\rightarrow \phi n$ cross section, which is taken from the
empirically measured one.  This approach, first introduced in Ref.
\cite{yao}, has been used earlier for calculating kaon production cross
section in the nucleon-nucleon interaction \cite{wuko,liko95b}. The  
results of Ref. \cite{sib} (dotted curve) are seen to be
significantly larger than our results, which
includes both the off-shell and interference effects. Experiments
on phi meson production from the proton-proton collision near the threshold 
are being carried out at SATURN \cite{kohn}. These data will be very useful 
in checking the validity of our model.

The extension of the model to phi production in 
nucleon-delta ($N\Delta$) and delta-delta ($\Delta\Delta$) interactions
is straightforward, except for the complication arising from the
fact that at the $\pi N\Delta$ vertex the energy momentum conservation 
allows the pion to go on-shell. A pole thus develops in the 
pion propagator at some region of the phase space, which then leads
to a singular cross section. Similar singularities appear also in 
$N\Delta \rightarrow N\Lambda K$ \cite{liko95b} and 
$N\Delta\rightarrow NN\eta$ \cite{peters}.
In both Ref. \cite{liko95b} and Ref. \cite{peters}, 
a complex pion self-energy has been introduced to remove the
singularity. Since the pion self-energy is density dependent, the
resulting cross section becomes density dependent, and 
diverges when the density approaches zero. This treatment
also introduces certain model dependence as different pion
self-energies have been used. In Ref. \cite{peters}, the pion self-energy is
related to the pion-nucleon scattering cross section, while in 
Ref. \cite{liko95b} it is calculated in a simple $\Delta$-hole model. 

In this work, this singularity is removed 
by using the so-called Peierls method \cite{peierls}, which
has also been used very recently in the treatment of 
the singularity in the reaction $\mu^{+} \mu^{-} \rightarrow 
e \bar \nu W^{+}$ associated with the physics of muon collider
\cite{muon}. The basic idea of the Peierls method is to
introduce a complex four-momentum for the resonance in the
initial state to account for its finite lifetime. As a result,
the four-momentum of the exchanged particle acquires an imaginary part
through the energy momentum conservation.
In our case, the pion propagator in \ndelnnphi and \deldelnnphi
is modified as follows:
\begin{eqnarray}
   \frac{1}{\left(p_{\Delta}-p_{N} \right)^{2} - m_{\pi}^{2} }
   \longrightarrow
   \frac{1}{\left(p_{\Delta}-p_{N} \right)^{2} - m_{\pi}^{2}
              -i m_{\Delta} \frac{(E_{\Delta}-E_{N})}{E_{\Delta}} 
               \Gamma_{\Delta} },
\end{eqnarray}
where $p_N$ and $p_\Delta$ are the four momenta of the 
final nucleon and the initial delta resonance, respectively; $E_N$ and 
$E_\Delta$ are their energies in the center-of-mass frame; and 
$m_\Delta$ is the delta mass.
The singularity at $(p_\Delta-p_N)^2=m_\pi^2$ is seen being removed by
the finite delta width $\Gamma_\Delta\approx 120$ MeV. The resulting cross 
sections are shown in Fig. 8.  Near threshold the \ndelnnphi cross section 
(dashed curve) is about a factor of five larger than that of \nnnnphi 
(solid curve), 
while the \deldelnnphi cross section (dotted curve) is only slightly larger
than the latter.  

In summary, we have calculated the phi meson production cross sections
in pion-baryon and baryon-baryon interactions based on a
boson-exchange model. Most parameters in the model 
are taken from the Bonn model for the nucleon-nucleon potential.
Two additional cut-off parameters are introduced at the $\pi\rho\phi$ vertex,
and are determined by fitting to available experimental data.
We have calculated the isospin averaged cross sections for 
\pinphin, \pidelnphi, \nnnnphi, \ndelnnphi, and \deldelnnphi
without introducing any further parameters. It is found that the cross
section for \pinphin is much larger than the other cross
cross sections for phi meson production. 
These cross sections will be useful for the transport model study of
phi meson production in heavy-ion collisions at SIS energies, and such a
study will be reported elsewhere \cite{chung}. Also, to study possible 
medium modifications of the phi meson properties, in particular the 
reduction of its mass, the measurement of dilepton spectra
will be very useful \cite{liko95a,wolf}.
Experimentally this will be carried out by the HADES collaboration 
for heavy-ion collisions at SIS energies\cite{hades}.
A detailed transport model study of dilepton production
including phi meson production from pion-baryon and
baryon-baryon interactions and from
other processes is also in progress.

\bigskip

We are grateful to W. Kohn for discussions, R. Machleidt 
and Z. Huang for communications,
and P. Lepage for making available to us the subroutine {\sc vegas}
for carrying out the numerical integrations
in the calculation of the elementary cross sections. 
This work was supported in part by
the National Science Foundation under Grant No. PHY-9509266. 
The support of CMK by the Alexander von Humboldt Foundation
is also gratefully acknowledged. GQL was also supported in part by
the Department of Energy under Grant No. DE-FG02-88ER40388.

\newpage

\pagestyle{empty}
\begin{figure}
\begin{center}
\vfill
\mbox{\epsfxsize=14truecm\epsffile{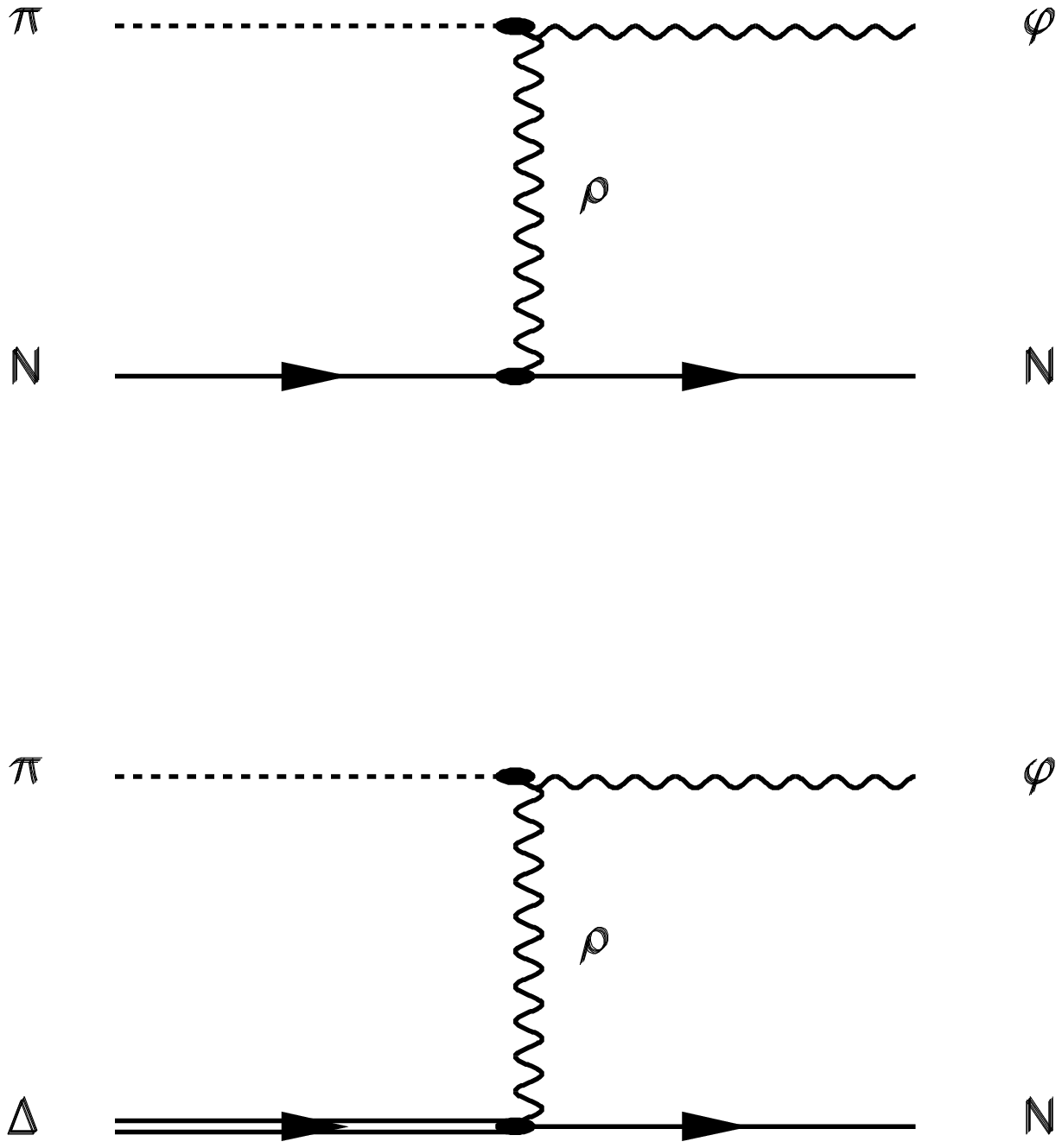}}
\caption{Feynman diagrams for \pinphin and \pidelnphi.
\label{npi} }
\vfill
\end{center}
\end{figure}

\pagestyle{empty}
\begin{figure}
\begin{center}
\vfill
\mbox{\epsfxsize=14truecm\epsffile{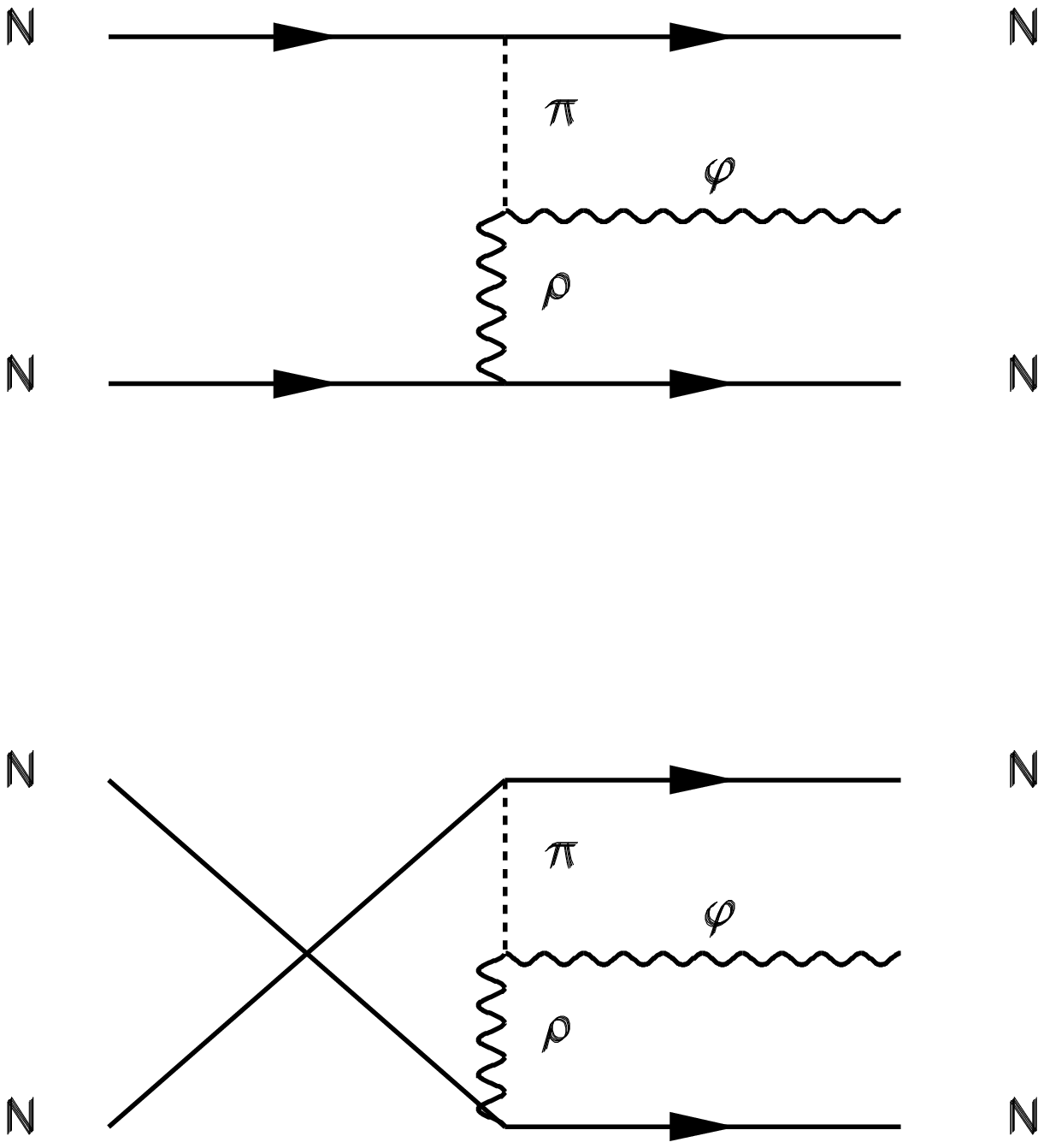}}
\caption{Feynman diagrams for \nnnnphi.
\label{nn} }
\vfill
\end{center}
\end{figure}

\begin{figure}
\begin{center}
\vfill
\mbox{\epsfxsize=14truecm\epsffile{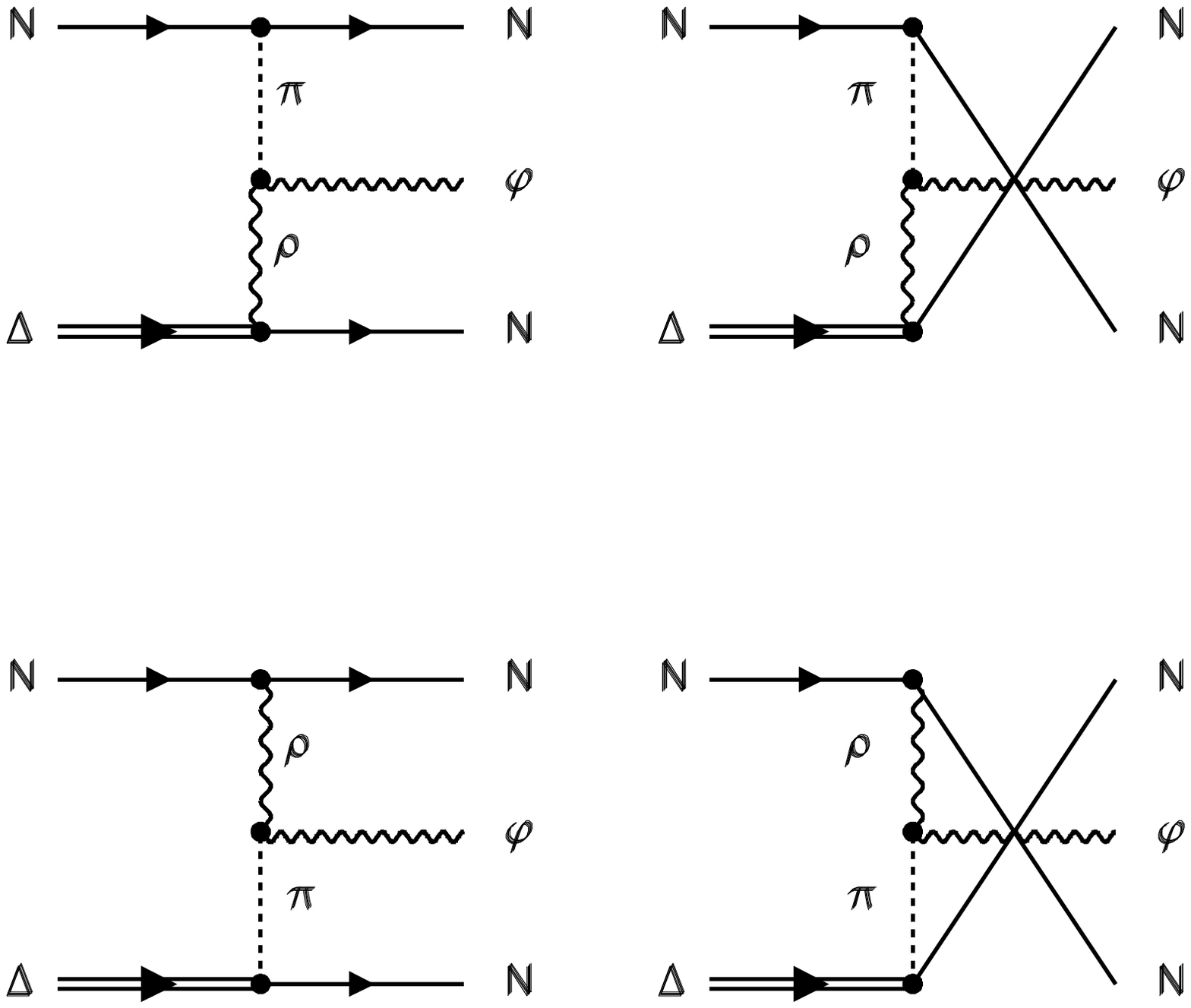}}
\caption{Feynman diagrams for \ndelnnphi.
\label{ndel} }
\vfill
\end{center}
\end{figure}

\begin{figure}
\begin{center}
\vfill
\mbox{\epsfxsize=14truecm\epsffile{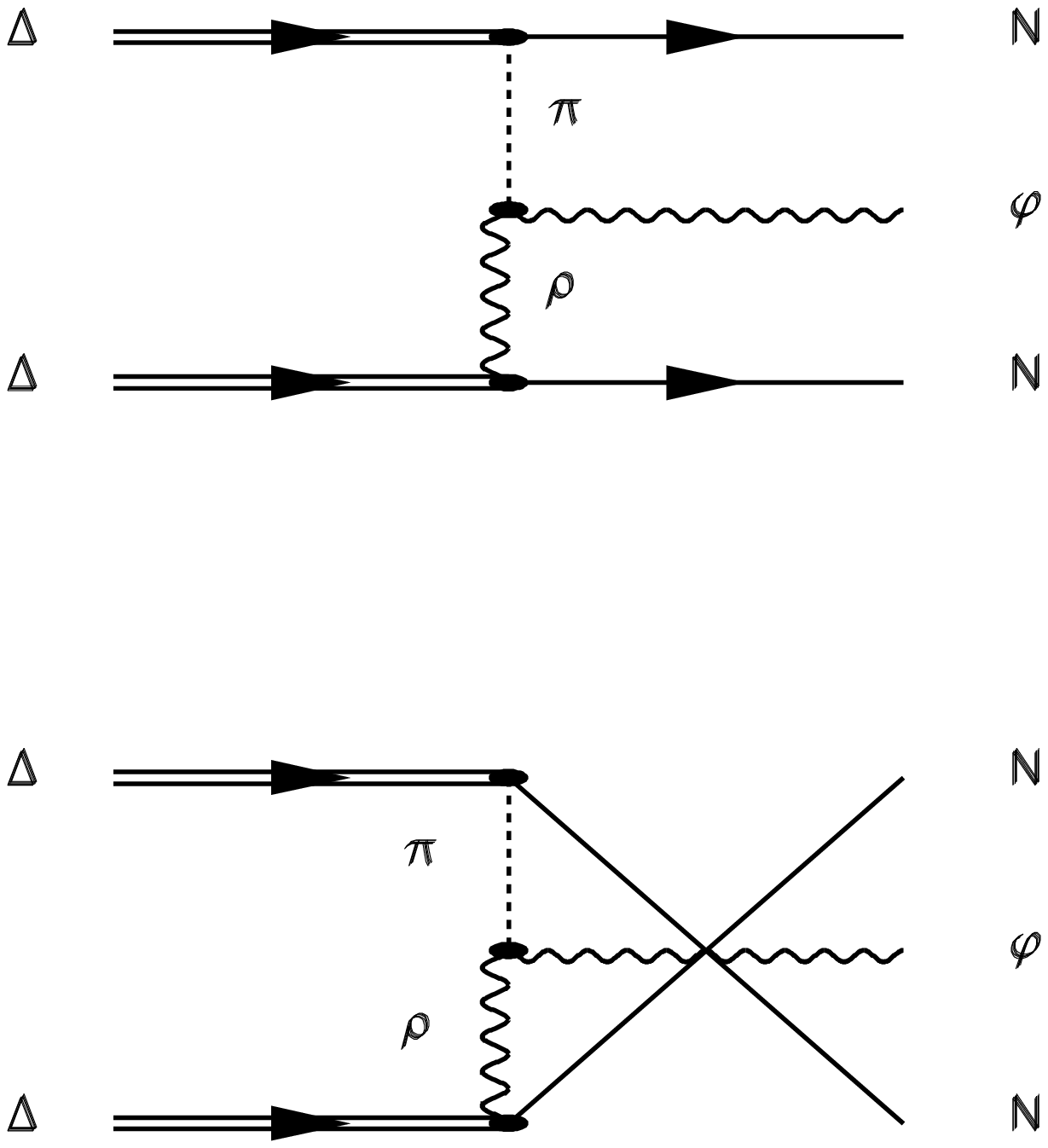}}
\caption{Feynman diagrams for \deldelnnphi.
\label{deldel} }
\vfill
\end{center}
\end{figure}

\begin{figure}
\begin{center}
\vfill
\mbox{\epsfxsize=14truecm\epsffile{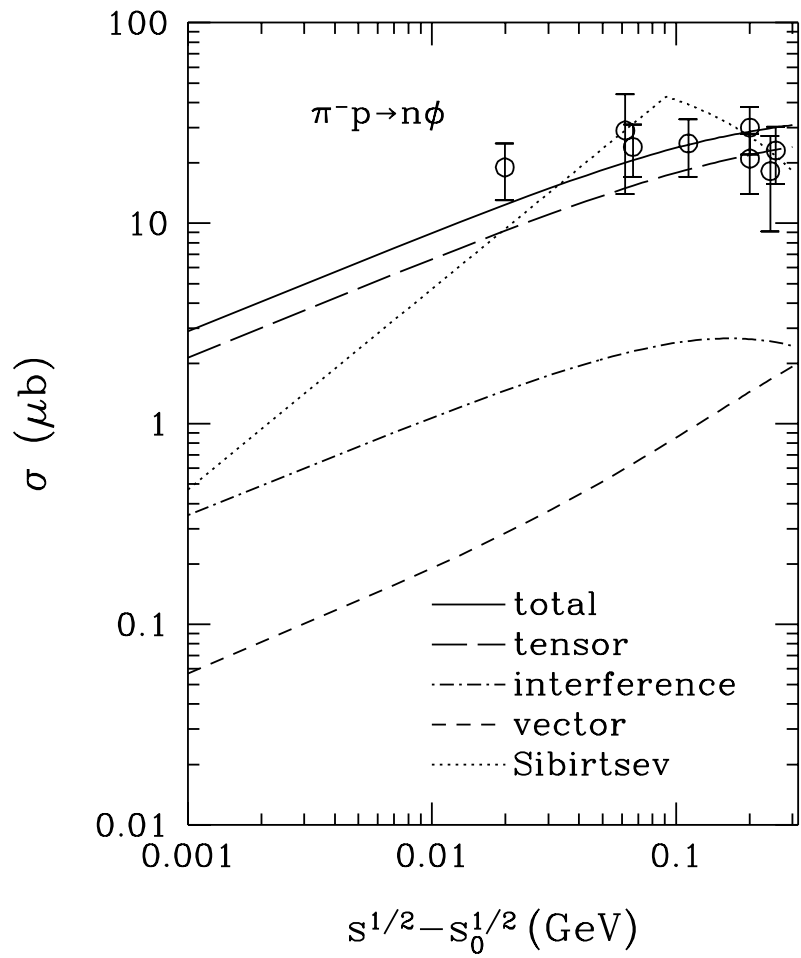}}
\caption{The cross section for $\pi^-p\rightarrow \phi n$ .
\label{piprot} }
\vfill
\end{center}
\end{figure}

\begin{figure}
\begin{center}
\vfill
\mbox{\epsfxsize=14truecm\epsffile{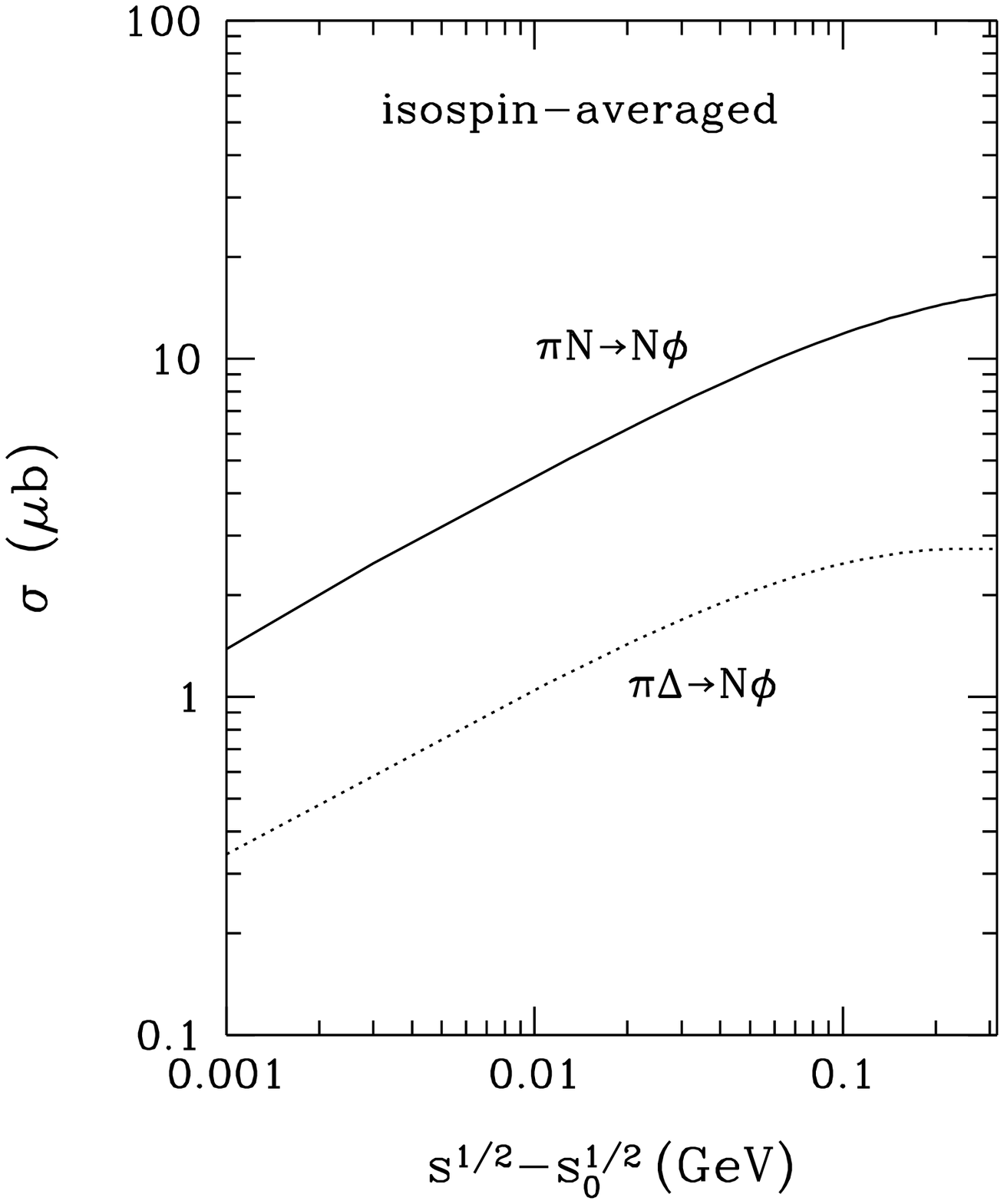}}
\caption{Isospin-averaged cross sections for 
\pinphin and \pidelnphi .
\label{pinpid} }
\vfill
\end{center}
\end{figure}

\begin{figure}
\begin{center}
\vfill
\mbox{\epsfxsize=14truecm\epsffile{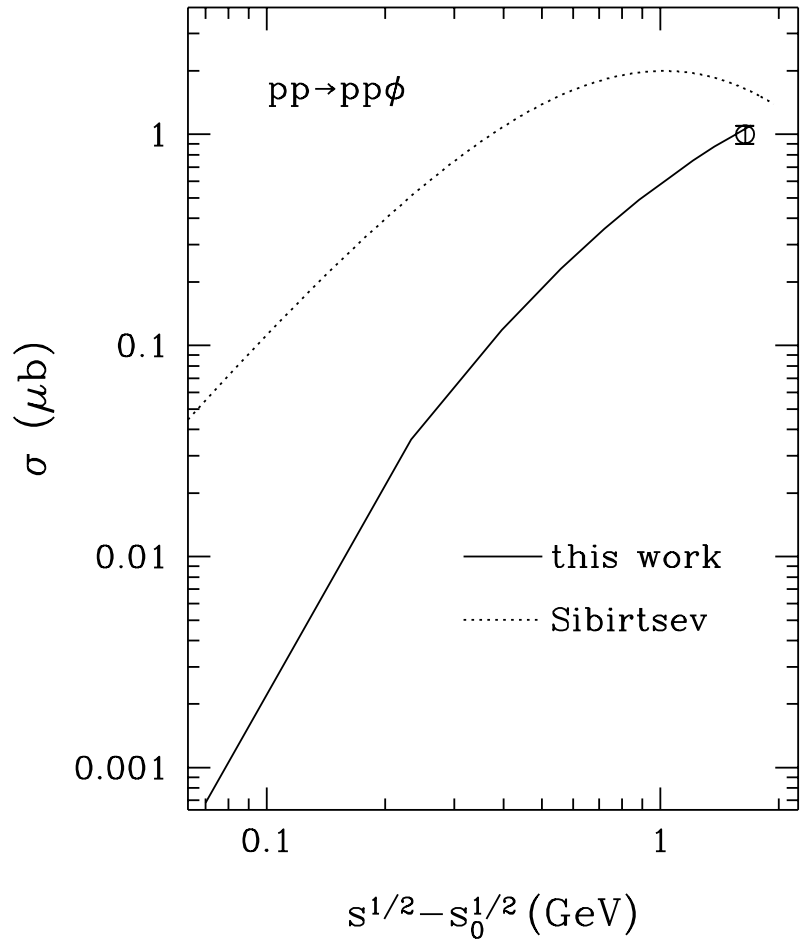}}
\caption{The cross section for $pp\rightarrow pp\phi$. 
\label{ppphi} }
\vfill
\end{center}
\end{figure}

\begin{figure}
\begin{center}
\vfill
\mbox{\epsfxsize=14truecm\epsffile{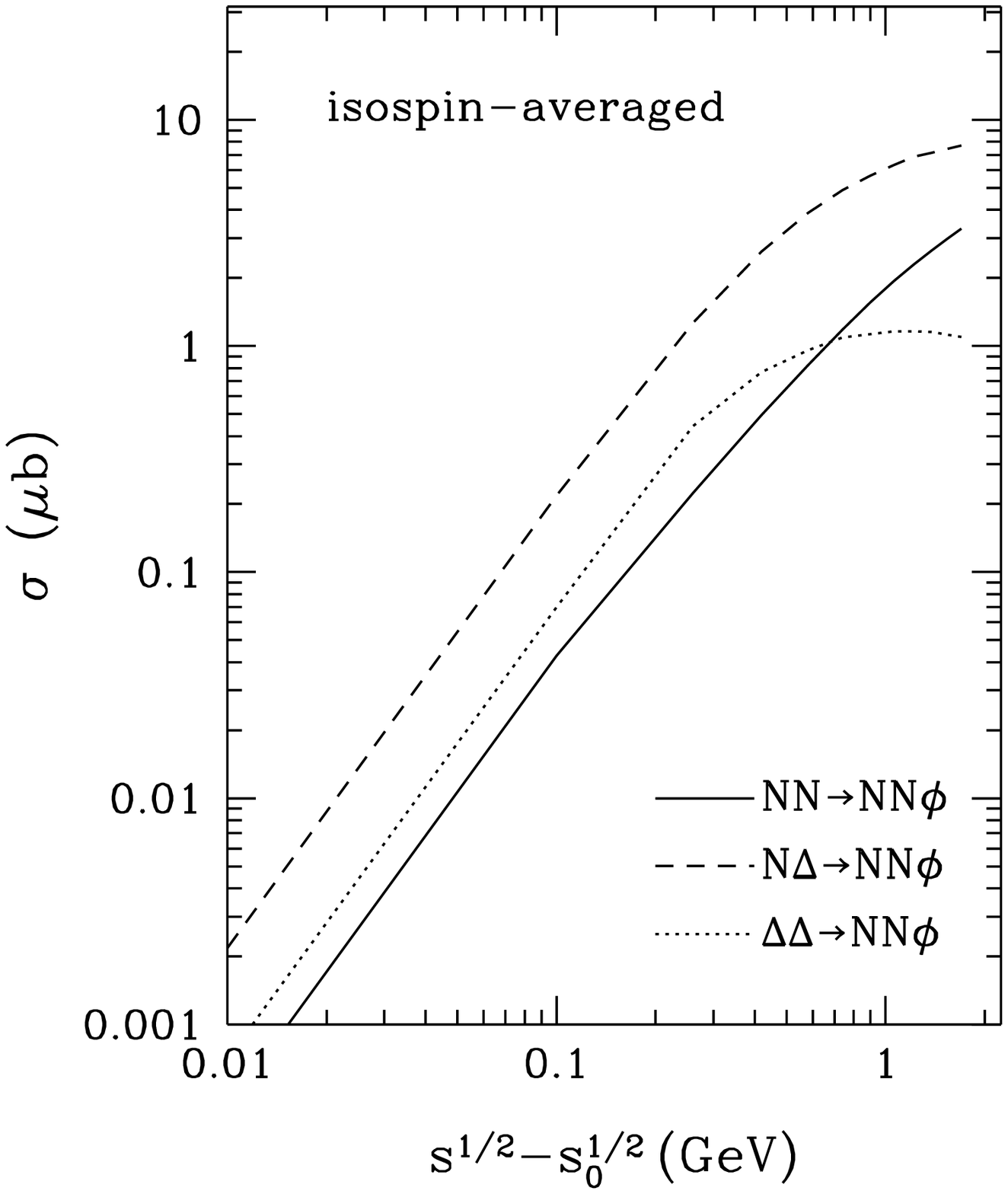}}
\caption{Isospin-averaged cross sections for \nnnnphi, \ndelnnphi and 
\deldelnnphi. \label{xsect} }
\vfill
\end{center}
\end{figure}


\begin{thebibliography}{99}
 
\bibitem{hols92} J. F. Donoghue, E. Golowich, and B. R. Holstein,
Dynamics of Standard Model, (Cambridge University Press, Cambridge, 1992).
 
\bibitem{shor85} A. Shor, Phys. Rev. Lett. 18 (1985) 1122.
 
\bibitem{hat92} T. Hatsuda and S. H. Lee, Phys. Rev. C 46 
(1992) R34.
 
\bibitem{ko94} M. Asakawa and C. M. Ko, Nucl. Phys. A 572 (1994) 732.
 
\bibitem{hat95} H. Kuwaraba and T. Hatsuda, Prog. Theor. Phys.
94 (1995) 1163.
 
\bibitem{song96} C. S. Song, Phys. Lett. B 388 (1996) 1410.

\bibitem{shur92} E. Shuryak and V. Thorsson, Nucl. Phys. A 536
(1992) 739.
 
\bibitem{ko91} C. M. Ko and B. H. Sa, Phys. Lett. B 258 (1991) 6.

\bibitem{na38}J. P. Guilland {\it et al.}, Nucl. Phys. A 525
(1991) 499c.
 
\bibitem{rqmd} M. Berenguer, H. Sorge, and W. Greiner, Phys. Lett.
B 332 (1994) 15.

\bibitem{koch90} P. Koch, U. Heinz, and J. Pis\'ut, Phys. Lett.
B 243 (1990) 149.
 
\bibitem{ags} Y. Akiba {\it et al.,}, Phys. Rev. Lett. 76 (1996) 2021.
 
\bibitem{fopi} N. Herrmann for FOPI Collaboration, in: Proc. Quark Matter
'96, Nucl. Phys. A610 (1996) 49c.

\bibitem{liko95a} G. Q. Li and C. M. Ko, Nucl. Phys. A 582 (1995) 731.
 
\bibitem{cass90} W. Cassing, V. Metag, U. Mosel, and K. Niita,
Phys. Rep. 188 (1990) 363.

\bibitem{mosel} U. Mosel, Ann. Rev. Nucl. Part. Sci. 41 (1991) 29.

\bibitem{koli96} C. M. Ko and G. Q. Li, J. of Phys. G 22 (1996) 1673.

\bibitem{eng} A. Engel, R. Shyam, U. Mosel, and A. K. Dutt-Mazumder,
Nucl. Phys. A603 (1996) 387.

\bibitem{wuko} J. Q. Wu and C. M. Ko, Nucl. Phys. A499 (1989) 810.

\bibitem{laget} J. M. Laget, Phys. Lett. B 259 (1991) 24.

\bibitem{liko95b} G. Q. Li and C. M. Ko, Nucl. Phys. A594 (1995) 439.

\bibitem{peters} W. Peters, U. Mosel, and A. Engel, Z. Phys. A 353
(1995) 333. 

\bibitem{grei} L. A. Winckelmann, H. Sorge, H. St\"ocker,
and W. Greiner, Phys. Rev. C51 (1995) R9.

\bibitem{mach89} R. Machleidt, Adv. Nucl. Phys. 19 (1989) 189.
 
\bibitem{meis88} U.-G. Meissner, Phys. Rep. 161 (1988) 213.

\bibitem{verm} J. A. M. Vermaseren, private communication. 

\bibitem{data1} A. Baldini {\it et al.,} Total cross sections of high
energy particles, (Springer-Verlag, Heidelberg, 1988).
 
\bibitem{sib} A. A. Sibirtsev, Nucl. Phys. A604 (1996) 455.

\bibitem{data2} R. Baldi {\it et al.,} Phys. Lett.
B68 (1977) 38.

\bibitem{data3} V. Blobel {\it et al.,} Phys. Lett. B59 (1975) 88.

\bibitem{yao} T. Yao, Phys. Rev. 125 (1961) 1048.

\bibitem{kohn} W. Kohn, private communication.

\bibitem{peierls} R. F. Peierls, Phys. Rev. Lett. 6 (1961) 641.

\bibitem{muon} I. F. Ginzburg, Nucl.  Phys. B (Proc. Suppl) 51A (1996) 85.

\bibitem{chung} W. S. Chung, G. Q. Li, and C. M. Ko, Nucl. Phys. A,
submitted.

\bibitem{wolf} Gy. Wolf, W. Cassing, and U Mosel, Nucl. Phys. A 552
(1993) 549.
 
\bibitem{hades} W. Koenig, in: Proc. Workshop on dilepton
production in relativistic heavy-ion collisions, ed. H. Bokemeyer
(GSI, Darmstadt, 1994).
 
\end{thebibliography}
\end{document}